# Understanding consumer demand for new transport technologies and services, and implications for the future of mobility

29 November 2018

Akshay Vij (corresponding author)
Institute for Choice
University of South Australia
Level 13, 140 Arthur Street
North Sydney, NSW 2060
vij.akshay@gmail.com
+61 08 8302 0817

**Abstract**

The transport sector is witnessing unprecedented levels of disruption. Privately owned cars that operate on internal combustion engines have been the dominant modes of passenger transport for much of the last century. However, recent advances in transport technologies and services, such as the development of autonomous vehicles, the emergence of shared mobility services, and the commercialization of alternative fuel vehicle technologies, promise to revolutionise how humans travel. The implications are profound: some have predicted the end of private car dependent Western societies, others have portended greater suburbanization than has ever been observed before. If transport systems are to fulfil current and future needs of different subpopulations, and satisfy short and long-term societal objectives, it is imperative that we comprehend the many factors that shape individual behaviour. This chapter introduces the technologies and services most likely to disrupt prevailing practices in the transport sector. We review past studies that have examined current and future demand for these new technologies and services, and their likely short and long-term impacts on extant mobility patterns. We conclude with a summary of what these new technologies and services might mean for the future of mobility.

## 1. Introduction

The transport sector has not seen this much disruption since Henry Ford invented the Model T in 1908. Privately owned cars that operate on internal combustion engines have been the dominant modes of passenger transport for much of the last century. Recent advances in transport technologies and services, such as the emergence of shared mobility services, the invention of connected and autonomous vehicles, and the commercialization of electric vehicle technologies, hold profound implications for future patterns of transport and land use behavior.

As Cortright (2016) writes, "The optimists see a world where parking spaces are beaten into plowshares, the carnage from car crashes is eliminated, where greenhouse gas emissions fall sharply and where the young, the old and the infirm, those who can't drive have easy access to door-to-door transit. The pessimists visualize a kind of exurban dystopia with mass unemployment for those who now make their living driving vehicles, and where cheap and comfortable autonomous vehicles facilitate a new wave of population decentralization and sprawl."

In the face of such uncertainty between these widely divergent scenarios, it becomes particularly salient that we understand how consumers will engage with these new systems and services, and what will be the consequent economic, social and environmental impacts of their decisions. Over subsequent sections, this chapter introduces the technologies and services most likely to disrupt prevailing practices in the transport sector, namely shared mobility services, connected and autonomous vehicles, and electric vehicles. We review past studies that have examined current and future demand for these new technologies and services, and their likely short and long-term impacts on extant mobility patterns. We follow this discussion with a review of other potential disruptors to the transport sector that might emerge in the future, such as unmanned aerial vehicles, 3D printing and hyperloop transport. Finally, we conclude this chapter with a summary of what these new technologies and services might mean for the future of mobility.

## 2. Shared mobility services

A number of recent 'megatrends' are disrupting the provision of transport services worldwide and reshaping the broader mobility landscape (JPI Urban Europe, 2017; Corwin et al., 2014). Major advances in information and communication technologies (ICTs) have created a digital economy, where new web-based services, such as e-commerce platforms, video messaging services, digital health services, online distance learning portals, etc., are changing the need and desire for travel (Cohen-Blankshtain and Rotem-Mindali, 2016). Many of these same advances have conspired to result in the emergence of new forms of shared mobility services, as represented by short-term carshare services such as ZipCar and GoGet, rideshare services such as Uber and Lyft, and public bikesharing services such as Capital Bikeshare and oBike, that are changing how consumers use the transportation system (Shaheen et al., 2017).

The rise of collaborative consumption and the growth in business and consumer interest in shared mobility services reflects a broader transition from an ownership-based economy to an access-based economy, particularly with regards to personal mobility (Belk, 2014). For example, in Australia alone, a country with a total population of roughly 24 million, carshare services have 66,000 members and offer access to 2,200 vehicles, rideshare services employ 72,000 active driver-partners and serve 3.3 million active ride seekers, and bikeshare services offer access to 7,000 bicycles nationwide. However, consumer interest has in some cases lagged behind: in a 2018 survey of 3985 Australians nationwide, 12.4 per cent reported using rideshare services a few times a month or more, but the corresponding numbers for carshare and bikeshare services were significantly lower at 5.6 per cent and 4.7 per cent, respectively (ITS Australia, 2018).

The transition away from private car ownership has been aided by concurrent economic and demographic shifts. The turn of the twenty-first century has seen suggestions of "peak car" (Goodwin and Van Dender, 2013), with stagnant or declining levels of private car use across much of the developed world, including Australia. For example, between 2001 and 2016, per capita vehicle kilometres travelled decreased by 6 per cent nationally, and licensing rates for people under 25 dropped by more than 10 per cent in Victoria and New South Wales. Studies have ascribed the apparent decline in private car dependence to a combination of economic factors, such as a recessionary global economy and rising oil prices, and demographic factors, such as an ageing population, rising higher education enrolment rates, an increase in the average age of entry into the labour market, and the decision to start a family at a later age (see, for example, Vij et al., 2017 and McDonald, 2015).

Over subsequent sections, we review rideshare, carshare and public bikeshare services. We introduce formal definitions for each of these services; we review their current and expected future status across different global markets; and we review consumer preferences for these different services.

## 2.1 Rideshare services

Ridesharing services refer to transportation network companies (TNCs) that use smartphone applications to match individuals wishing to make a trip from a specified origin to a specified destination with individuals willing to drive them there in their personal cars (Rayle et al., 2014). Like taxis, ridesharing services too offer point-to-point transportation. However, there are some key differences between the two: (1) individuals wishing to request or provide rides must register with the ridesharing service before they can use it; (2) rides are crowdsourced from a pool of available drivers that consists largely of part-timers, usually not licensed to drive commercial vehicles, looking to supplement their incomes from other jobs; (3) the ridesharing service employs location-based smartphone technology and data mining algorithms to reduce waiting times, increase reliability and adjust fares in real-time; (4) payments for each trip are processed online using billing information provided at the time of registration by the individuals requesting and providing the ride; and (5) after the trip, the individual who requested the ride can leave feedback about the individual who provided the ride, and this information is visible to other users of the service.

These differences independently may appear marginal at best, but together they have helped create a new paradigm for transportation. Uber, DiDi and Ola Cabs have emerged perhaps as the preeminent ridesharing services currently operating in the world: Uber offers ridesharing services in 633 cities worldwide; DiDi in 400 cities across China; and Ola Cabs in 110 cities in India. Both DiDi and Ola Cabs have plans for expansion into overseas markets, with the latter having recently launched in Australia. Traditional taxi and bus services are adopting some of the same service-based principles championed by ridesharing services. For example, ongoing on-demand public transport trials in Newcastle, New South Wales offer potential passengers the option of booking services through a smartphone app. Similarly, apps like GoCatch and InGoGo allow users to make taxi bookings through smartphone interfaces that are similar to those used by ridesharing services. Increasingly, the boundaries between ridesharing, taxis and public transport services are becoming less clear.

The competition between traditional taxi services and newer ridesharing services has prompted several academic enquiries. Most studies agree that ridesharing services have had the greatest impact on existing taxi services. For example, Nelson (2016) finds that the annual number of taxi trips in Los Angeles declined from 8.4 million in 2013 to 6.0 million in 2015. Similarly, Hu (2017) reports that, "for the first time, more people are using Uber in New York than the city's fabled yellow cabs. In July [2017], Uber recorded an average of 289,000 rides each day compared with 277,000 taxi trips." Some studies have found that ridesharing services have also been used to substitute trips that would otherwise have been made using public transport services or privately-owned cars. For example, Rayle et al. (2016) find in their survey of 380 intercepted rideshare users in San Francisco that at least half of the ridesharing trips were replacing modes other than taxi, including public transport and driving.

By and large, studies have found that users of rideshare services tend to be young, well-educated, high-income, employed individuals, with low levels of car ownership, living in dense urban environments (e.g. Dias et al., 2017; Rayle, 2016). Studies that have examined consumer preferences for different rideshare service attributes are rarer in the literature. In their analysis of 3,985 Australians nationwide, Vij et al. (2018) find that while consumers are willing to pay, on average, 0.28$/km more to avoid sharing a vehicle with other passengers and 0.17$/km more for door-to-door service, cost is the most important determinant of rideshare use. For a rideshare service that costs roughly the same as UberX's ridesharing service ($1.15 per km), and offers comparable level-of-service, the study predicts that 17 per cent of the national population could be expected to use the service a few times a week or more. However, if the same service could be provided at a much lower cost of $0.30 per km, through potential advances in electric, connected and autonomous vehicle technologies, a significantly larger 31 per cent of the national population could be expected to use the service a few times a week or more.

## 2.2 Carshare services

Carshare services are short-term car rental services that offer consumers access to a private car when and where they need one, without the costs associated with ownership or maintenance. While carsharing has existed in different forms since the earliest days of the automobile, it has only become widely available as a mode of transport since 2000, enabled in large part by the internet (for a comprehensive discussion on the origins of carsharing, the reader is referred to Shaheen and Cohen, 2013).

Carshare services may offer access to cars through both peer to peer (P2P) and business to consumer (B2C) models. The P2P model matches individuals wishing to rent their privately-owned cars with individuals needing short-term access to one, with the sharing system operated by a third-party. The B2C model is similar to traditional car rental companies, where a single organization offers customers access to a fleet of vehicles owned by the organization itself at one or more stations, but differs in the following critical ways: (1) customers wishing to access cars must register with the carshare service, and in most cases they must also pay a monthly or yearly subscription cost to have access to the service; (2) vehicles may be rented by the minute, the hour or the day, depending on the service, and fuel costs are inclusive; (3) vehicles are typically distributed across the service area at popular points of departure, either at fixed stations or free floating within designated areas; and (4) customers can usually access and return cars any time of day, and they can track vehicles and make reservations in real time. Carshare service operators might offer roundtrip services, where customers must pick-up and drop-off cars at the same location, or one-way services, where customers may pick-up and drop-off cars at different locations.

The size of the global carsharing market was estimated to be USD 1.5 billion in 2017, and is expected to grow to USD 11 billion by 2024 (Global Market Insights, Inc., 2018). In North America alone, as of 2017, B2C carsharing services had roughly 1.9 million members and a combined fleet of 24,629 vehicles across thirty-nine operators (Shaheen and Cohen, 2017), and P2P carsharing services had roughly 2.9 million participating individuals and a combined fleet of 131,336 vehicles across six operators (Shaheen et al., 2018a). Future growth is expected to be led by newly industrializing countries like China and India. The Chinese B2C carsharing market already has a combined fleet of 26,000 vehicles, and is projected to grow at 45 per cent per annum until 2025 (Roland Berger, 2018).

Notwithstanding this growth, profitability continues to be a concern globally, particularly for B2C carshare service operators with high capital and operating costs, and recent years have seen a number of carshare service operators withdraw from specific markets (e.g. Deschamps, 2018; Jackson, 2017). Many national and regional governments have offered support to carshare service operators to help them sustain operations locally. For example, most Chinese operators still "rely on government subsidies and are still not running profitable and sustainable business models" (Roland Berger, 2018). In Australia, support from regional and local governments has been lacking in many cases, and some have blamed this limited support for low use (Phillip Boyle & Associates, 2016).

Carshare members tend to be young and highly educated, frequently university students or white-collar professionals, and often part of moderate-income non-traditional households (see, for example, Becker et al., 2017; Efthymiou and Antoniou, 2016; Schmöller et al., 2015; Cervero et al., 2007). Additionally, carshare members are likely to have low levels of car ownership and high levels of public transport patronage, leading studies to conclude that carsharing services act as substitutes to private car ownership and complements to public transport. For example, in a survey of 6,281 carshare members in North America, Martin et al. (2010) find that 25 per cent of their sample reduced their level of car ownership after membership, and an additional 25 per cent postponed their decision to purchase a car. Based on further analysis, the study concludes that a single shared car has the capacity to replace between 9 and 13 privately owned cars, and a more recent Australian study corroborates this finding (AECOM, 2016). In their analysis of carshare users in Toronto, Canada, Costain et al. (2012) find that carsharing is most often used for off-peak period travel or on weekends, when public transport service is poor. Their findings are echoed by similar studies conducted in Switzerland (Becker et al., 2017a), Italy (De Luca and Di Pace, 2015) and the United States (Zoepf and Keith, 2016) that find carsharing services are often used by customers to plug gaps in existing public transport services.

In terms of service attributes, one-way services are found to be more popular than round-trip services, often by a factor of three to four (see, for example, Schmöller et al., 2015 and Le Vine et al., 2014). Carsharing costs, and how they compare with alternative modes of travel, are found to be strong determinants of use, as evidenced by findings from North America (Burkhardt and Miller-Ball, 2006), Europe (De Luca and Di Pace, 2015) and Asia (Yoon et al., 2017). Guaranteed access to a vehicle is found to have a strong positive effect on membership (Kim et al., 2017), and the number of vehicles in the carshare service network is found to increase frequency of use (Habib et al., 2012). Surprisingly, most studies do not find access distance to carshare vehicles to be a significant determinant of carshare use (Yoon et al., 2017), and "that customers are willing to accept a substantially longer access walk to the car-sharing vehicle than for public transportation" (Becker et al. 2017b). Similarly, in their analysis of members of a North American carshare service, Zoepf and Keith (2016) find that customers are willing to pay only USD 2 per hour more in terms of rental costs to reduce access distance by one mile. However, most of these studies surveyed existing carshare users, and it is very likely that individuals who currently do not use carshare services have very different preferences.

## 2.3 Bikeshare services

Bikeshare services are bike rental companies that offer customers short-term access to bicycles. While bikeshare services have been around since the 1960s, like other shared mobility services, their popularity too has really surged in the last decade due to advances in ICTs. For a recent comprehensive review of the academic literature on bikeshare programs across the world, the reader is referred to Fishman (2016).

In terms of service design, bikeshare services share many features with carsharing services: (1) customers wishing to access bicycles must register with the service, and in some cases monthly or yearly subscription costs and initial registration fees might apply; (2) bicycles may be rented by the minute, the hour or the day, depending on the service; (3) bicycles are typically distributed across the service area at popular points of departure, either at fixed stations, often referred to as 'docks', or fully free floating, resulting in the more recent model of dockless bikeshare services; and (4) customers can usually access and return bicycles any time of day, and they can track available bicycles and make reservations in real time.

The number of cities with bikeshare services has grown to over two thousand, and most major metropolitan regions in the world currently have one or more bikeshare services in operation (Meddin and DeMaio, 2018). The number of bicycles available through these services worldwide has increased commensurately, from 700,000 in 2013 to 2.3 million in 2016 (Bernard, 2018). However, again like carsharing services, profitability continues to be a concern, and many bikeshare service operators have been compelled in recent months to close operations (Tchebotarev, 2017). In many cities, the public sector has actively supported bikeshare services, either by taking on the role of service provider through government-run operations, or more frequently, through public-private partnerships.

Greater public sector involvement has typically been motivated through the benefits that increased bicycling can offer, in terms of its impacts on car use and traffic congestion on one hand, and population health outcomes on the other (Fishman, 2016). For example, in their analysis of bikeshare programs across five cities in Australia, Europe and the US, Fishman et al. (2014b) find that the programs reduced car use in four of the five cities. However, their analysis also finds that bikeshare services are more frequently used to substitute public transport and walking, than they are to substitute driving, and their finding is supported by other studies as well (e.g. Zhu et al., 2013). In terms of impacts on health and physical activity, in their analysis of bikeshare users in London, Woodcock et al. (2014) find that mean physical activity increased by an average of 0.06 MET hours (or 0.06 kcal per kg in bodyweight) per week per person as a result of joining the bikeshare service. As the authors argue, "although this is small on average at the individual level, it led to notable modelled gains in health at population level."

Bikeshare users tend to have higher incomes (Fishman et al., 2015; Woodcock et al., 2014; Lewis, 2011), are more educated (Fishman et al., 2014a; LDA Consulting, 2013; Shaheen et al., 2013), and more likely to be employed (Woodcock et al., 2014). Some studies have also reported gender and racial differences between bikeshare users and general populations, with bikeshare users being more likely to be male and white (Fishman et al., 2014; Goodman and Cheshire, 2014; Buck et al., 2013;).

Barriers to greater use of bikeshare services centre primarily around concerns for safety, and relatedly, mandatory helmet laws, though some studies have also highlighted lengthy registration processes as additional impediments (Fishman et al., 2016). Despite these concerns, there is no clear agreement on the impacts of bikeshare services on road safety and traffic accidents. For example, in their analysis of hospital injury data from five US cities with bikeshare services and five without, over a three year period that extended two years before the service first started and one year after, Graves et al. (2014) find that the proportion of head injuries among bicycle-related injuries increased in cities with bikeshare services, from 42 per cent before the service to 50 per cent after, leading the authors to conclude that "steps should be taken to make helmets available with PBSPs [public bicycle share programs]." However, Cowling (2014) and Salomon et al. (2014) have contested the validity of these conclusions. Based on their own analysis of the same data, the two studies find that annual *total* bicycle injury rates decreased by 28 per cent in the cities with bikeshare services (despite the increase in the *proportion* of head injuries), leading them to conclude "that overall bike safety improves with PBSPs, possibly because of increased driver awareness or improved biking infrastructure" (Cowling, 2014).

## 3. Connected and autonomous vehicles

Connected vehicles are vehicles that use ICTs to communicate with the driver, other road users, roadside infrastructure and other wireless services. Autonomous vehicles are vehicles where one or more primary driving controls, such as steering, acceleration and braking, do not require human input for sustained periods of time. Together, connected and autonomous vehicles (CAVs) have the capacity to offer a number of social benefits that include increased road safety, higher traffic flows, greater travel time productivity, improved energy efficiency, greater accessibility, etc. The technology is currently being trialled all across the world, including Australia, both on public roads and more controlled, 'closed-loop' conditions, such as university campuses and retirement villages. The first commercially available fully autonomous car is expected to be available by 2020.

Many of these CAV technologies will likely be offered to potential consumers as both products and services. For example, car companies such as Tesla and Ford are planning to integrate automated features within existing car models. Concurrently, carsharing and ridesharing companies such as Uber are investing in these technologies with the intention of integrating them within existing services. It is anticipated that CAV technologies will enable on-demand door-to-door transport services as a new form of micro public transport, which combines the benefits of existing mass public transport services and private modes of motorized transport, but does not suffer from the same drawbacks (Wong et al., 2017). Compared to mass public transport services that require large catchment areas in order to be feasible, and consequently suffer from first and last mile connectivity problems, micro public transport can offer door-to-door services. Compared to private modes of motorized transport, where high parking costs and frequent congestion can limit access and use, micro public transport is expected to be cheaper, faster and more convenient. Therefore, any analysis of the potential impacts of CAVs must necessarily account for ongoing and future competition between ownership-based and sharing-based models of mobility.

Numerous studies have sought to understand public perceptions of CAVs (Menon et al., 2016; Duncan et al., 2015; Kyriakidis et al., 2015; Payre et al., 2014; Rodel et al., 2014; Schoettle and Sivak, 2014; Casley et al., 2013; Howard and Dai, 2013). Commonly identified perceived benefits include greater safety, better fuel economy, and more productive use of travel time. And commonly identified concerns include equipment and system failure, cyber security, data privacy, legal liability in case of crash, and loss of control.

From an ownership standpoint, on average, studies find that consumers are willing to pay roughly $3,000 for partial automation, and roughly $5000 – $7500 for complete automation (e.g. Daziano et al., 2017; Bansal et al., 2016). However, as mentioned previously, shared CAVs could erode current consumer willingness to pay for private CAV ownership. While private ownership will still likely appeal to niche segments, such as families with young children, tradespeople with heavy equipment, etc., shared CAVs could help a significant proportion of the general population transition from owning two cars to one car, and potentially even to no cars. On average, privately owned cars are not in use 95 per cent of the time (Shoup, 2017). With automation, it can be expected that many of these cars will likely be available for short-term rental through P2P carshare services, further diminishing the need for private ownership.

Consequently, traditional car manufacturers are looking to replace potential revenue lost because of reduced car sales by taking on the role of transport service providers themselves. For example, BMW already operates carsharing services in North America and Europe, and General Motors plans to commence its own rideshare services in 2019 using self-driving cars developed in house. Simultaneously, existing shared mobility service providers such as Uber, in addition to investing heavily in the development of CAV technology, are continuing to subsidize their current services, with a long-term view towards holding on to their present advantage within the point-to-point transport market.

In light of these developments, several studies in recent years have examined the latent demand for shared CAV services. Krueger et al. (2016), in their survey of 435 Australians nationwide, find that service attributes such as travel time, waiting time and fares will be significant determinants of consumer adoption of shared CAV services, and young travellers will likely be the early adopters. Bansal et al. (2016), in their survey of 347 residents of Austin, Texas in the United States, find that only 13 per cent of survey participants would be willing to give up personal vehicles and rely exclusively on shared CAVs that cost roughly $1/mile, and at least 35 per cent of survey participants would be unwilling to use shared CAV services at all, regardless of their costs. Haboucha et al. (2017), in their survey of 721 individuals living

across Israel and North America, find that consumers are still hesitant to embrace CAV technology, and that even if shared CAV services were completely free, 25 per cent of their sample would still be unwilling to use the service. Hao and Yamamoto (2017), in their case study on Meito Ward, Nagoya in Japan, predict that up to 30 per cent of total trips conducted in the region could be served by shared CAVs in the future.

Given both the uncertainty that still surrounds CAVs (in terms of the technology itself, the supporting infrastructure, and the regulatory framework) and consumer unfamiliarity (consumer surveys have repeatedly found that significant proportions of the general population are unfamiliar with CAV technology and how it will likely function; see, for example, Schoettle and Sivak, 2014), any research on the potential demand for private CAV ownership and shared CAV use has had to be, by necessity, somewhat speculative. Predicted adoption rates from these studies will likely change as the technology matures and as consumers become more familiar with corresponding services.

## 4. Electric vehicles

Electric vehicles (EVs) have existed for more than a hundred years, and were among the earliest cars available to the general public. In 1900, prior to the emergence and subsequent dominance of internal combustion engines, EVs comprised roughly one-third of the American car fleet (DoE, 2014). Rising oil prices and climate change concerns have fuelled a twenty-first century resurgence. It began with the introduction of Toyota Prius in 1997, a hybrid electric vehicle (HEV) that has an electric drive system and battery, but cannot be plugged in to the electric grid. This was followed by the creation of Tesla Motors in 2003 with the explicit objective of accelerating consumer adoption of EVs. Tesla's creation spurred other car manufacturers to develop their own EVs. In 2010, General Motors launched the Chevrolet Volt, a plug-in hybrid electric vehicle (PHEV) which, like HEVs, has a gasoline engine that supplements its electric drive once the battery is depleted, but unlike HEVs, can be plugged in to the electric grid to recharge the battery. That same year, Nissan launched its Leaf model, a battery electric vehicle (BEV) that is all-electric, does not depend on petrol and produces no tailpipe emissions. Two years later, in 2012, Tesla launched its Model S, also a BEV. Today, there are 35 PHEV and 36 BEV models available on the global market.

Notwithstanding this interest from industry in the production of EVs, consumer demand has been slower to catch up. High costs, low driving ranges, long charging times and limited public charging infrastructure continue to be major impediments to adoption, though most experts agree that the electrification of our transport infrastructure is not a question of if, but when (for a comprehensive recent review of consumer preferences for EVs, the reader is referred to Liao et al., 2017). In 2017, BEVs and PHEVs comprised only 1.7 per cent of total passenger car sales worldwide. However, while total sales remain low, year-on-year growth has averaged 30-40 per cent in the last five years, and this rate of growth is expected to continue over coming years (Scutt, 2018).

The growth in EV sales has been led by burgeoning demand in China and parts of Europe. In both cases, government support has been essential to stimulating and sustaining consumer demand. In the case of China, government support has been motivated in terms of industrial development. The Chinese government has taken an active interest in positioning China at the forefront of EV research and development activities, and a large local market could help Chinese car manufacturers progress along the global value chain. The Chinese government wants EVs to account for 12 per cent of total car sales by 2020, compared to 3.7 per cent in 2017 (Lee, 2018). In its attempt to achieve this ambitious target, the government has been offering subsidies to EV buyers of up to 110,000 yuan (or USD 16,000) per unit. To promote local development of EVs with higher driving ranges, subsidies have been set to be higher for EVs with greater driving ranges (Dixon, 2018). Additionally, the government plans to have 500,000 public charging stations nationwide by 2020, up from roughly 200,000 at the end of 2017 (Fusheng, 2018), to further support consumer adoption.

In the case of more developed European nations, government support has been motivated by concerns around climate change and oil dependence. Norway in particular has led the way in consumer adoption. The country has the highest market penetration per capita in the world, with roughly 5 per cent of all vehicles on Norwegian roads being PHEVs or BEVs (Manthey, 2018), EVs comprised a remarkably high 39 per cent of total car sales in 2017 (Knudsen and Doyle, 2018). Local adoption has been helped by high fuel prices, cheap electricity, and most significantly, generous government incentives. Until 2017, EVs were exempt from "value added tax and purchase tax, which on average in Norway add 50% to the cost of a vehicle. They are also exempt from road tolls, tunnel-use charges, and ferry charges. And they get free parking, free charging, and the freedom to use bus lanes" (Mirani, 2015). However, national and local governments have indicated that some of this support will be withdrawn over coming years, as the market has matured and subsidies and incentives are no longer deemed necessary to sustain further growth (Nelson, 2017).

## 5. Other potential disruptions to the transport sector

The coming years could see additional disruptions in the form of unmanned aerial vehicles (UAVs), 3D printing, and hyperloop transport. Most of these technologies are still in their research and development phase, with few, if any, commercial applications to date.

UAVs, or drones, are perhaps the most developed of these new technologies. Large drones could potentially displace existing long-distance air, rail and freight shipping industries. Companies like Natilus plan to start testing drones with capacities of up to 100 tons by 2020. Commercialization is still likely at least 5-10 years away, but if and when the technology is market ready, it could reduce long-haul shipping costs by as much as half (Terdiman, 2017). Smaller drones could similarly displace urban delivery services. Several companies working in the space of logistics, retail and food, such as United Parcel Service (UPS), Amazon and Domino's, are currently testing drone delivery systems. Regulators and policy-makers in North America and Europe are reacting to these developments accordingly. For example, the Federal Aviation Authority (FAA) in the United States recently launched its Unmanned Aircraft System (UAS) Integration Pilot Program (IPP) as "an opportunity for state, local, and tribal governments to partner with private sector entities, such as UAS operators or manufacturers, to accelerate safe UAS integration". Studies that have examined public perceptions of drones report that privacy and security concerns will likely be major barriers to consumer adoption (see, for example, Chang et al., 2017).

3D printers have become increasingly commercially available in the last decade, but mainstream adoption is yet to happen. Unlike current manufacturing processes that rely on high production volumes to achieve optimal economies of scale, 3D printers could potentially usher in a more distributed and small-scale approach to manufacturing. As Lim and Nair (2017) write, "The advent of 3D printing opens the way for manufacturers to significantly reduce the production cost of their goods by eliminating many steps in the manufacturing process, such as casting and welding metal. It also reduces the complete production process to no more than three to four key players. With 3D printing, what would have initially been a series of stages of production could be cut down to a designer at one end, and the printer or "manufacturer" at the other. The middle players would most likely be suppliers of raw materials or 'ink'… Such reductions in the manufacturing process could affect both regional and international production networks, possibly resulting in reduced capital requirements, warehousing and other logistics and transportation needs." In essence, 3D printing could shorten supply chains by allowing goods to be manufactured closer to the end consumer, and therefore, reduce delivery distances of products (Shaheen et al., 2018b).

A hyperloop is a sealed tube or system of tubes through which a pod may travel free of air resistance or friction conveying people or objects at high speed (Musk, 2013). Hyperloop transport could do for long-distance inter-city travel what CAVs promise to do for short-distance urban and regional travel. For example, Musk's original paper conceives "a hyperloop system that would propel passengers along the 350-mile (560 km) route at a speed of 760 mph (1,200 km/h), allowing for a travel time of 35 minutes, which is considerably faster than current rail or air travel times". However, most experts express scepticism about the technology and its purported potential (see, for example, Taylor et al., 2016), and in the absence of real-world implementations, it is hard to anticipate what impacts, if any, the technology might portend for the future of mobility.

## 6. Preparing for the future of mobility

The future of passenger transport is predicted to be electric, autonomous, and increasingly shared (Firnkorn and Müller, 2015). Each of these changes by themselves would be potentially disruptive; together they are expected to be transformative, with far-reaching economic, social and environmental ramifications. For example, a 2015 report by the Committee for Economic Development of Australia (CEDA) on Australia's future workforce estimated that 3.28 million Australians employed nationally had jobs that involved driving. "If we assume that over the next 20 years all vehicles become autonomous, these job categories will need to change into something completely different as the human skill of driving a vehicle is no longer the essential part of the job" (CEDA, 2015). Electric vehicles could lead to significant reductions in oil consumption and GHG emissions, contributing to the creation of a more sustainable transport system. However, they will require significantly more energy than the electricity grid can currently provide, and provision of appropriate public charging infrastructure to support consumer adoption. Shared mobility services have already been credited with reduced private car ownership and use in some parts of the developed world (Martin et al., 2010). In conjunction with CAVs, they could help accelerate an ongoing transition from two-car households to one-car households, and potentially even to zero-car households, for at least a subset of the population. However, they could also potentially increase net vehicle kilometres travelled, due to greater suburbanization as enabled by reduced travel costs.

If as engineers, planners and policy-makers, we are to influence the process to produce societally optimal outcomes, we must better comprehend the determinants of individual behaviour. In the past, both industry and government have failed to anticipate the rate and scale of change with regards to similarly new technologies and services. For example, the rapid diffusion of rideshare services such as Uber in Australia has caught both taxi providers and transport planners by surprise. State governments have been forced to introduce expensive bailouts to keep local taxi industries afloat and compensate taxi plate owners for their losses. Legislation has had to be amended to allow rideshare companies like Uber to operate legally, following pressure from consumer advocacy groups. New regulations are under review to address concerns around unfair employment practices, inadequate insurance coverage, and lax security measures with regards to the provision of these services. Some of this disruption could have been avoided had the public and private sectors been appropriately prepared.

The impacts of connected, autonomous and electric vehicles, and the accompanying disruption, will only be greater, but we seem to be making the same mistakes. Consider, for example, the ongoing Sydney light rail (SLR) project. Initial plans were announced in 2011 – as a way to connect Sydney's eastern suburbs with the urban centre. The service is planned to commence operations in 2019 and is projected to transport 9.7 million customers every year. However, these projections do not account for shared CAVs which could emerge in the near future as a feasible new form of public transport. From an engineering standpoint, the SLR might be the most efficient means of transport in terms of throughput (i.e. number of passengers moved through a corridor). However, from a consumer standpoint, mass public transport services like the SLR require large catchment areas in order to be feasible, and consequently suffer from first and last mile connectivity problems. In contrast, shared CAVs could enable on-demand door-to-door public transport services that are likely to be much more attractive to customers. The SLR business case does not account for potential loss in demand due to these new technologies and services (TfNSW, 2013), which could seriously undermine the project's long-term economic viability and profitability, and render the project obsolete well in advance of planned end-life dates.

The case of ridesharing serves as an excellent example for why we need better tools for understanding consumer response to new transport technologies and services; modelling and forecasting their consequent impacts; and identifying ways in which industry and government can best prepare themselves for these imminent transformations, in an attempt to produce outcomes that are economically efficient, socially equitable and environmentally sustainable.